\documentclass[prl,twocolumn,floatfix,showpacs,tightenlines]{revtex4}
\usepackage{graphicx}

\begin{document}

\title{Steady-state nonequilibrium density of states of driven strongly
correlated lattice models in infinite dimensions} 
\author{A.~V.~Joura$^*$ and J.~K.~Freericks$^*$}
\affiliation{$^*$Department of Physics, Georgetown University, 37th and O Sts. NW, Washington, DC
20057, U.S.A.},
\author{Th.~Pruschke$^\dagger$}
\affiliation{$^\dagger$Institute for Theoretical Physics, University of G\"ottingen, Friedrich-Hund-Platz 1,
D-37077 G\"ottingen, Germany}

\date{\today}

\begin{abstract}
The formalism for exactly calculating the retarded and advanced Green's
functions of strongly correlated lattice models in a uniform electric field
is derived within dynamical mean-field theory.  To illustrate the method, we solve for
the nonequilibrium density of states of the Hubbard model in both the
metallic and Mott insulating phases at half-filling (with an arbitrary
strength electric field) by employing the numerical renormalization group as
the impurity solver.  This general approach can be applied to any strongly
correlated lattice model in the limit of large dimensions.
\end{abstract}

\pacs{71.27.+a, 71.10.Fd, 71.45.Gm, 72.20.Ht}

\maketitle

{\it Introduction}. The many-body formalism for nonequilibrium problems
was formulated independently by Kadanoff and Baym~\cite{kadanoff_baym_1962} and 
Keldysh~\cite{keldysh_1965} in the 1960s. One of the main applications of that
work was to determine the nonlinear transport properties of strongly
correlated materials.  Recently there has been a significant emphasis placed
on examining small open systems (quantum dots attached to leads) within the 
Meir-Wingreen~\cite{meir_wingreen} generalization of the Kadanoff-Baym-Keldysh
approach and there has been progress in applying
Bethe ansatz techniques to this problem~\cite{andrei}, numerical renormalization group techniques within a scattering state formalism~\cite{anders}, and Hirsch-Fye quantum Monte Carlo techniques by mapping to an effective imaginary time formalism~\cite{han}. That work is timely
because the size of local electric fields placed over nanoscale electronics
is enormous and nonlinear quantum effects are likely to be critical for
understanding how those systems behave. In this work, our focus is
on larger systems (bulk materials) placed under large electric fields,
which serves as a counterpart (top-down) approach to the problem, and
could have direct application in ultracold atomic systems placed in
optical lattices and driven into nonequilibrium by accelerating the
lattice through space (by detuning the frequencies of the counterpropagating 
lasers, or simply from the force of gravity).

There has also been much effort applied to understanding the original
Kadanoff-Baym-Keldysh formalism and how to approximately solve the
resulting equations.  The generalized Kadanoff-Baym approximation~\cite{gkba}
and the reconstruction theorem for the lesser Green's 
function~\cite{reconstruction} have provided much insight into the way
quantum systems relax and ultimately reach a steady state. But there remains no
exact solutions for strongly correlated
bulk systems placed in large electric fields in
the steady state.  In this contribution, we partially solve this problem
by showing how to calculate the retarded and advanced Green's functions,
and hence the many-body density of states (DOS) within the dynamical
mean-field theory approach~\cite{metzner_vollhardt_1989}. 
Since these Green's functions are a needed input
into the reconstruction theorem, this can be viewed as an initial step
toward a complete steady-state formalism. The formal development here is
more general than the transient response formalism~\cite{noneq_prl} because
it can be applied to any many-body lattice Hamiltonian that can be solved 
with a real time or real frequency impurity solver in a dynamical mean field;
here we show results for the Hubbard model solved with the numerical 
renormalization group (NRG) method.

{\it Formalism}. Our focus is on the advanced and retarded Green's functions.
Since the advanced Green's function is directly related to the retarded
Green's function via complex conjugation and an interchange of the two time
variables, we will derive results for the retarded Green's function only.
We use the Keldysh boundary condition for the nonequilibrium problem:
starting our system in an equilibrium distribution at a constant temperature
and then turning on the constant and spatially uniform
electric field.  We then let the system 
evolve forward in time until all transients have died off and we
are left with the steady-state response. The electric field is described by
a spatially uniform vector potential in the Hamiltonian gauge, where the 
scalar potential vanishes ${\bf E}(t)=-\partial {\bf A}(t)/c\partial t$
and we ignore all magnetic field effects that are present near the time the 
field is initially turned on.  For a uniform field, we then have ${\bf A}(t)=-c
{\bf E}t$ since the field is turned on in the infinite past (but after the
system has reached equilibrium at temperature $1/\beta$). The vector potential
is input into the Hamiltonian via the Peierls' substitution~\cite{peierls_1933}, so that the
nonequilibrium Hamiltonian is translationally invariant and can be
described in momentum space.  The momentum-dependent retarded
Green's function is defined to be
\begin{equation}
G^R_{\bf k\sigma}(t,t^\prime)=-i\theta(t-t^\prime){\rm Tr}
e^{-\beta \mathcal{H}_{eq}}
\{c^{}_{\bf k\sigma}(t),c^\dagger_{\bf k\sigma}(t^\prime)\}_+/\mathcal{Z}_{eq},
\label{eq: gr_def}
\end{equation}
where the averages are taken with respect to the initial equilibrium 
Hamiltonian ($\mathcal{Z}_{eq}$ is the equilibrium partition function), and 
the time evolution of the creation ($c^\dagger_{\bf k\sigma}$) and
annihilation ($c^{}_{\bf k\sigma}$) operators (for electrons with momentum
{\bf k} and spin $\sigma$) is in the Heisenberg picture. We will examine the
case where the electric field lies along the diagonal of a hypercubic
lattice in $d$ dimensions ${\bf E}=E(1,1,1,...)$ and then the limit
where $d\rightarrow\infty$.

The retarded Green's function and self-energy $\Sigma^R$
satisfy a Dyson equation that
is formally equivalent to the equilibrium Dyson equation except that all
functions now depend on two time variables instead of just the time difference
\begin{eqnarray}
G_{\bf k\sigma}^R(t,t^\prime)&=&G_{\bf k\sigma}^{R0}(t,t^\prime)
\label{eq: dyson_time}\\
&+&\int d\bar t \int d\bar t^\prime G_{\bf k\sigma}^{R0}(t,\bar t)
\Sigma^R_{\sigma}(\bar t,
\bar t^\prime)G_{\bf k\sigma}^R(\bar t^\prime,t^\prime),
\nonumber
\end{eqnarray}
where the noninteracting Green's function can be found 
exactly~\cite{turkowski_freericks_2005,book}
\begin{eqnarray}
&~&G_{\bf k\sigma}^{R0}(t,t^\prime)=-i\theta(t-t^\prime)\exp \left [
-i\int_{t^\prime}^t d\bar t (\epsilon_{{\bf k}+E\bar t}-\mu)\right ],
\nonumber\\
&~&G_{\bf k\sigma}^{R0}(T,t_{rel})
= -i\theta(t_{rel})e^{i\mu t_{rel}}\nonumber\\
&\times&\exp \left [ -i\frac{2(\epsilon_{\bf k}\cos ET-
\bar\epsilon_{\bf k}\sin ET)}{E}\sin \frac{Et_{rel}}{2}\right ].
\label{eq: gr0_def}
\end{eqnarray}
Here, the bandstructure $\epsilon_{\bf k}$ 
satisfies~\cite{metzner_vollhardt_1989} 
$\epsilon_{\bf k}=
-\lim_{d\rightarrow\infty}t^*\sum_{i=1}^d\cos {\bf k}_i/\sqrt{d}$,
the second bandstructure $\bar\epsilon_{\bf k}$ satisfies
$\bar \epsilon_{\bf k}=
-\lim_{d\rightarrow\infty}t^*\sum_{i=1}^d\sin {\bf k}_i/\sqrt{d}$ (all
energies will be measured in units of $t^*$ and we set $c=1$), 
and the second line uses the
Wigner coordinates of average time $T=(t+t^\prime)/2$ and relative time
$t_{rel}=t-t^\prime$.
We will always work with the paramagnetic solution, so we can drop the
spin label on all functions, since the system will be symmetric between 
spin-up and spin-down, and there is no long-range magnetic order.

The self-energy has no momentum dependence because we are working in the
infinite-dimensional limit; the perturbative result of 
Metzner~\cite{metzner_1991} and the Langreth rules~\cite{langreth} show
that the self-energy remains local in nonequilibrium as well. But working in the
steady state does provide some additional simplifications to the time
dependence which needs to be exploited to be able to solve the problem.
To start, note that the noninteracting Green's function satisfies the
{\it gauge property} which relates shifts in momentum to shifts in average time
\begin{equation}
G^{R0}_{{\bf k}+{\bf E}\bar t}(T,t_{rel})=
G^{R0}_{\bf k}(T+\bar t,t_{rel}),
\label{eq: gauge}
\end{equation}
where we write the Green's function as a function of the Wigner coordinates.
The second property is the {\it Bloch periodicity property}, which shows the
system is periodic in the steady state
\begin{equation}
G^{R0}_{\bf k}(T+2\pi/E,t_{rel})=
G^{R0}_{\bf k}(T,t_{rel}).
\label{eq: bloch}
\end{equation}
The gauge property implies that the local noninteracting retarded 
Green's function
is independent of average time, and the Bloch periodicity property
implies that the momentum-dependent
noninteracting retarded Green's function is periodic in average time with
the Bloch period $2\pi/E$.  Note that it is incorrect to assume that the
steady state has no average time dependence.  Indeed, the noninteracting
steady-state momentum-dependent
Green's functions explicitly depend on average time.

The next step requires an ansatz that we cannot yet prove to be rigorously true.  The ansatz is
that the local retarded self-energy also has no average time dependence. The argument
supporting this is that we start the dynamical mean-field theory algorithm
with a self-energy equal to zero.  Then the local Green's function is equal
to the noninteracting Green's function and is average time independent, 
hence the effective medium is average time independent and so is the
impurity Green's function.  Then the new self-energy will also be average
time independent.  Continuing in this fashion, we see that the DMFT algorithm
will not introduce any average time dependence into the problem.  What we 
cannot rule out is that there may exist solutions with an average time
dependent self-energy; in this case we would not have any simple criterion 
to choose which one is the correct solution to the nonequilibrium problem.  
When we analyzed the Falicov-Kimball model~\cite{noneq_prl} with a
transient formalism, we saw that the transient solution does approach the
steady-state solution at long times, which is an example showing that in
at least some cases the self-energy does evolve to an average time independent
self-energy at long times~\cite{book}. If the self-energy has no average time
dependence, then the dressed retarded Green's function satisfies both the
gauge property and the Bloch property, and in particular, the local dressed
retarded Green's function is independent of average time.

Using these results, we will perform a continuous
Fourier transformation of all functions
with respect to relative time, and a discrete Fourier series expansion
with respect to average time (with Fourier frequencies $\nu_n=nE$, integer
multiples of the Bloch frequency) [$G_{\bf k}^R(T,t_{rel})=\sum_n\int d\omega 
G_{\bf k}^R(\nu_n,\omega) \exp(-i\nu_nT-i\omega t_{rel})/2\pi$]. The 
momentum-dependent Dyson equation in Eq.~(\ref{eq: dyson_time}) becomes
\begin{eqnarray}
G_{\bf k}^R(\nu_n,\omega)&=&G_{\bf k}^{R0}(\nu_n,\omega)
\label{eq: dyson_fourier}\\
&+&
\sum_{m}G_{\bf k}^{R0}(\nu_m,\omega+{\textstyle\frac{1}{2}}\nu_n-
{\textstyle\frac{1}{2}}\nu_m)\nonumber \\
&\times&
\Sigma^R(\omega+{\textstyle\frac{1}{2}}\nu_n-\nu_m)
G_{\bf k}^R(\nu_n-\nu_m,\omega-{\textstyle\frac{1}{2}}\nu_m),
\nonumber
\end{eqnarray}
which couples together the Green's functions at frequencies differing by 
multiples of the Bloch frequency; this equation has an underlying matrix
structure to it that allows it to be solved in a straightforward fashion.
In particular, we can restrict $0\le \omega <E$ when solving the equation,
and determine the Green's function at all $\omega+\nu_n$.

Now we describe the generalization of the iterative DMFT 
algorithm~\cite{jarrell_1992}
for the nonequilibrium steady-state problem:
(i) begin with a guess for the electronic self-energy (usually chosen to be
zero); (ii) calculate the local retarded Green's function by solving
Eq.~(\ref{eq: dyson_fourier}) for each momentum point and summing over momentum
(which requires a two-dimensional integration over $\epsilon$ and $\bar\epsilon$
with a joint DOS $\rho(\epsilon,\bar\epsilon)=\exp(-\epsilon^2-\bar\epsilon^2)/
\pi$~\cite{turkowski_freericks_2005}); (iii) extract the effective 
medium for the impurity
problem from the Dyson equation for the impurity, using the local Green's
function and the old self-energy; (iv) solve the impurity problem in the
extracted effective medium to find the new impurity Green's function;
(v) use the impurity Dyson equation with the new impurity Green's function and 
the old effective medium to extract the new self-energy; and (vi) repeat
steps (ii)--(v) until the equations have converged. Since the impurity
problem depends only on the relative time, we can perform a Fourier 
transformation to real frequencies and use an impurity solver like the NRG
to solve the impurity problem for the impurity self-energy.  All of the
nonequilibrium effects enter through the momentum-dependent Dyson equation
and the self-consistency condition.  Once the retarded Green's function is
known, we compute the interacting DOS from $\rho(\omega)=
-{\rm Im}G^R_{loc}(\nu_n=0,\omega)/\pi$.
Once the DOS has been determined, we can compute the first three moment sum
rules~\cite{moment} and use them to test the overall accuracy of the 
computation along with the zeroth and first self-energy sum rule~\cite{moment2}; we
work at half-filling, so particle-hole symmetry guarantees that the odd moments vanish.
In all cases considered here, the zeroth moment of the DOS satisfied the sum rule
to 1\% or better, the second moment of the DOS had errors in the range from 1--25\%,
while the zeroth moment of the self-energy had errors in the 15--30\% range.  All of these
results are better than what is typically seen in equilibrium calculations.

\begin{figure}[htb]
\centerline{\includegraphics [width=3.3in, angle=0]  {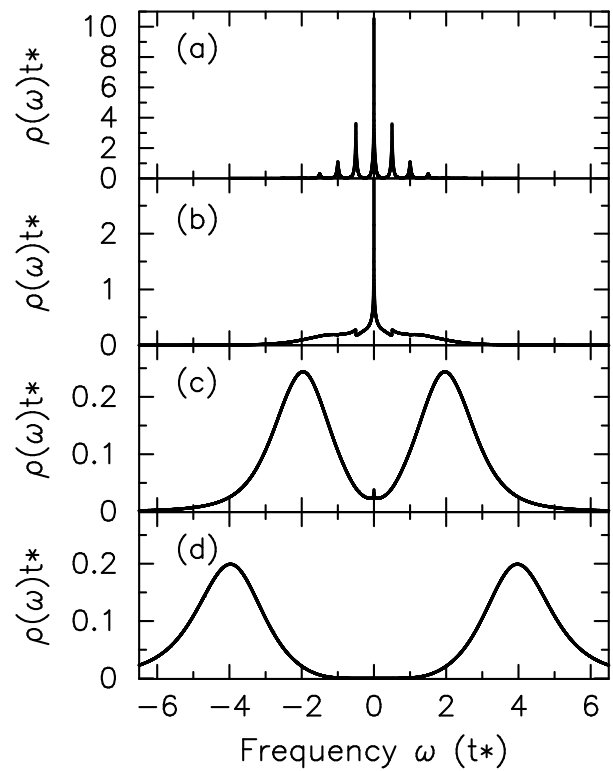}}
\caption[]{
Density of states for the infinite-dimensional Hubbard model with $E=0.5$.
The panels run from systems that are metallic to insulating (when in equilibrium): (a) $U=0.5$;
(b) $U=2$; (c) $U=4$; and (d) $U=8$. Note the change in the vertical axis size for the different panels.
}
\label{fig: e=0.5dos}
\end{figure}


\begin{figure}[htb]
\centerline{\includegraphics [width=3.3in, angle=0]  {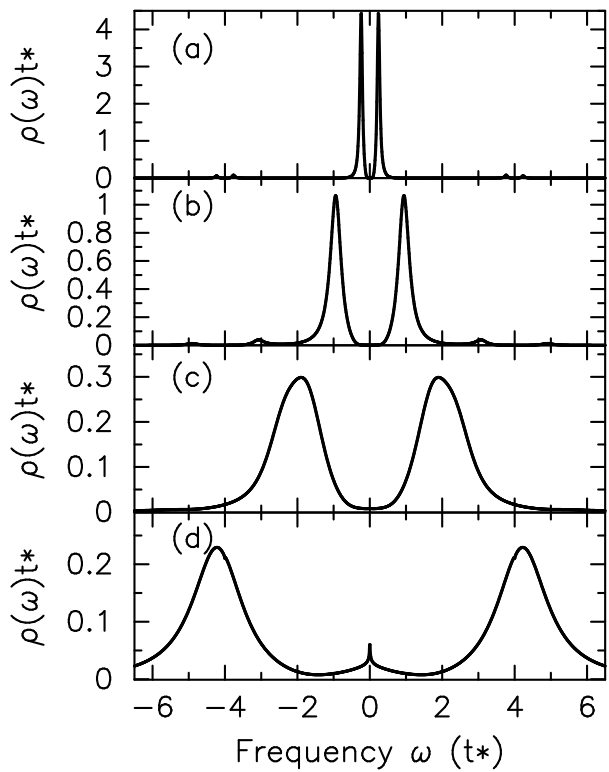}}
\caption[]{
Density of states for the infinite-dimensional Hubbard model with $E=4$.
The panels run from systems that are metallic to insulating (when in equilibrium): (a) $U=0.5$;
(b) $U=2$; (c) $U=4$; and (d) $U=8$. Note the change in the vertical axis size for the different panels.
}
\label{fig: e=4dos}
\end{figure}

{\it Results}. For concreteness, we solve for the nonequilibrium steady-state
response of the Hubbard model~\cite{hubbard} in infinite dimensions; the
Hubbard model involves electrons hopping between nearest-neighbor sites,
with an on-site repulsion $U$.  The equilibrium Hamiltonian is
\begin{equation}
\mathcal{H}_{eq}=\sum_{\bf k\sigma}\epsilon_{\bf k}c^\dagger_{\bf k\sigma}
c^{}_{\bf k\sigma}+U\sum_{\bf k,q,p}c^\dagger_{\bf k\uparrow}
c^{}_{\bf k-q\uparrow}c^\dagger_{\bf p\downarrow}c^{}_{\bf p+q\downarrow},
\label{eq: hubbard}
\end{equation}
and the nonequilibrium Hamiltonian results from the Peierls' 
substitution ($\epsilon_{\bf k}\rightarrow
\epsilon_{{\bf k}+{\bf E}t}$ for the steady-state problem).

Details of the NRG algorithm appear in Ref.~\cite{nrg_rev}.  In most cases,
we take $\Lambda=1.6$ and keep 1600 states. 
We begin our results by showing calculations for a weak field case, where
$E=0.5$ in Fig.~\ref{fig: e=0.5dos}. The four panels show progressively
larger values of the interaction strength $U$ ranging from metals (in equilibrium)
to Mott insulators.  In panel (a), we have the weak coupling result with $U=0.5$.  This behaves as expected, 
showing a broadening of the Wannier-Stark ladder delta functions, which are located at integer multiples of the 
Bloch frequency ($0.5n$ here). As the interactions increase further, the minibands broaden and merge into
a single band, but with a shape unlike that seen in equilibrium [panel (b) for $U=2$], and then they
start to form the upper and lower Hubbard bands [panel (c) for $U=4$ and panel (d) for $U=8$]; note that in panel (c) there is still a small peak appearing at the center of the density of states. The evolution with $U$ can be easily understood.
When the interactions are small enough, the driving field $E$ primarily determines the DOS and we have a broadened Wannier-Stark ladder.  As the broadening increases, the minibands merge into a single band.  Increasing the interactions further brings in Mott physics which dominates the behavior and the DOS approaches the shape of the equilibrium DOS.

Note that the peak at low frequency should not be confused with the quasiparticle peak that appears in equilibrium. Indeed, the self-energy does not have a Fermi liquid form at all.  The imaginary part does approach zero at $\omega=0$ for small $U$, but it does not have a quadratic behavior in $\omega$, and instead looks more like a cusp.  When the interaction strength gets large enough, the self-energy develops a sharp peak at $\omega=0$ reminiscent of the delta function peak for the Mott insulator in equilibrium.


In Fig.~\ref{fig: e=4dos}, we plot the nonequilibrium DOS for the strong field case of $E=4$.  Here the Bloch frequencies occur at $4n$.  In weak coupling, the DOS corresponds to the Wannier-Stark ladder with the delta functions broadened and now split by $U$ [panel (a) with $U=0.5$]; most of the spectral weight lies around $\omega=0$, but there are still visible peaks around $\omega=\pm 4$.  When $U$ is increased, the minibands also merge as in panel (b) for $U=2$.  One can see the splitting of the delta functions continues to increase, producing structure at $4n\pm 1$ now. In panel (c), we see a DOS for $U=4$ that looks quite similar to the equilibrium DOS (but at a nonzero $T$).  Panel (d) shows a new, and interesting effect.  This case, with $U=8$, has the split delta functions from the first bands at $\omega=\pm 4$ ``meeting'' at $\omega=0$.  The system responds by creating a small weight quasiparticle-like peak in the DOS at $\omega=0$.  Although this looks like a quasiparticle peak, the self-energy does not behave like a Fermi liquid here, so the origin is different.  Indeed, we find that the self-energy has a three-peak structure for small $U$---two broad peaks centered near $\omega=\pm U$ and a narrow peak centered at $\omega=0$.  As $U$ is increased, the broad self-energy peaks remain at $\omega\approx \pm U$, but the low-frequency peak disappears.  In this regime, the self-energy looks like it is developing a power-law cusp at low frequency.  The case with $U=8$ is anomalous, with the shape of the DOS differing significantly from what is seen at $U=6$ or $U=10$. 

{\it Conclusions}. In this work, we have shown how to generalize DMFT to nonequilibrium steady state situations, which can be mapped onto the same form of impurity problem that one uses to solve problems in equilibrium.  We are able to explicitly determine the retarded and advanced Green's functions, which we illustrate by examining the Hubbard model driven by different magnitude electric fields. We find a rich array of behavior, including a broadening of the Wannier-Stark ladders, an evolution toward the equilibrium DOS when $U$ is large enough, and a splitting of the Wannier-Stark delta functions when $E$ is large.  The self-energy also is anomalous, and never appears to illustrate behavior similar to that of a Fermi liquid---the nonequilibrium steady state simply behaves differently.

{\it Acknowledgments}.  We acknowledge useful conversations with
A. Hewson, A.-P. Jauho, V. S. Oudovenko, J. Serene, V. M. Turkowski, and V. Zlati\'c.  
This work is supported by the N. S. F. under grant number DMR-0705266.

\end{document}